\documentclass{aastex}
\usepackage{natbib}
\usepackage{emulateapj5}
\slugcomment{To appear in {\it The Astrophysical Journal, Letters}}
\shorttitle{Limits on FUV Emission}
\shortauthors{Dixon, et al.}

\begin{document}


\newcommand{\cf}{cf.}
\newcommand{\eg}{e.g.}
\newcommand{\etal}{et~al.}
\newcommand{\etc}{etc.}
\newcommand{\fig}[1]{Fig.~\ref{#1}}
\newcommand{\h}{$^{\rm h}$}
\newcommand{\ie}{i.e.}
\newcommand{\m}{$^{\rm m}$}

\newcommand{\cfour}{\ion{C}{4}}
\newcommand{\ctwo}{\ion{C}{2}}
\newcommand{\ebv}{$E($\bv)}
\newcommand{\f}{{\it f}\/}
\newcommand{\hone}{\ion{H}{1}}
\newcommand{\htwo}{H$_2$}
\newcommand{\kms}{km s$^{-1}$}
\newcommand{\lu}{photons cm$^{-2}$ s$^{-1}$ sr$^{-1}$}
\newcommand{\nho}{$N$(\hone)}
\newcommand{\nht}{$N$(H$_2$)}
\newcommand{\oone}{\ion{O}{1}}
\newcommand{\osix}{\ion{O}{6}}
\newcommand{\rv}{$R_V$}
\newcommand{\specfit}{{\small SPECFIT}}
\newcommand{\surface}{photons cm$^{-2}$ s$^{-1}$ \AA $^{-1}$ sr$^{-1}$}

\newcommand{\chips}{{\it CHIPS}}
\newcommand{\cop}{{\it Copernicus}}
\newcommand{\euve}{{\it EUVE}}
\newcommand{\fuse}{{\it FUSE}}
\newcommand{\hst}{{\it HST}}
\newcommand{\hut}{HUT}
\newcommand{\iue}{{\it IUE}}
\newcommand{\orf}{{ORFEUS}}
\newcommand{\rosat}{{\it ROSAT}}

\newcommand{\na}{New~Astronomy}

\title{FUSE Limits on FUV Emission from Warm Gas in Clusters of
Galaxies\footnotemark}

\footnotetext[1]{Based on observations made with the NASA-CNES-CSA {\it
Far Ultraviolet Spectroscopic Explorer.  FUSE} is operated for NASA by
the Johns Hopkins University under NASA contract NAS5-32985.}

\author{W.\ Van Dyke Dixon\altaffilmark{2}, Shauna Sallmen, and Mark Hurwitz}
\affil{Space Sciences Laboratory\\
University of California, Berkeley, CA 94720-7450, USA}

\email{vand, sallmen, markh@ssl.berkeley.edu}

\altaffiltext{2}{Current address: Department of Physics and Astronomy,
The Johns Hopkins University, Baltimore, MD 21218, USA}

\and

\author{Richard Lieu}
\affil{Department of Physics\\
University of Alabama in Huntsville, Huntsville, AL 35899, USA}

\email{lieur@cspar.uah.edu}

\begin{abstract}

We have obtained FUV spectra of two clusters of galaxies with \fuse,
the {\it Far Ultraviolet Spectroscopic Explorer}.  The Coma cluster was
observed for a total of 28.6 ksec, the Virgo cluster for 10.9 ksec.
Neither spectrum shows significant \osix\ $\lambda \lambda 1032, 1038$
emission at the cluster redshift.  Such emission would be expected from
the warm [$(5-10) \times 10^5$ K] component of the intracluster medium
(ICM) that has been proposed to explain the excess EUV and SXR flux
present in {\it EUVE} and {\it ROSAT} observations of these clusters.
Our 2-$\sigma$ upper limits on the \osix\ $\lambda 1032$ flux from Coma
and Virgo exclude all published warm-gas models of the EUV excess in
these clusters.

\end{abstract}


\section{INTRODUCTION}

Recent observations with the {\it Extreme Ultraviolet Explorer (EUVE)}
provide evidence that a number of clusters of galaxies are strong
emitters of EUV radiation \citep{Lieu:96a, Lieu:96b, BLM:97, MLL:98}.
This emission is substantially brighter than would be expected from the
well-known, X-ray--emitting intracluster medium (ICM;
\citealt{Forman:Jones:82}).  Signatures of this ``soft excess'' are
sometimes present in the lowest energy-resolution band of the
\rosat\ PSPC, but its presence remains controversial
\citep{Arabadjis:99, Drake:99, Krick:00}.  At present, two groups agree
that an EUV excess exists in both the Virgo and Coma clusters, but they
differ on the intensity and morphology of the emitting regions
\citep{LIAB:99, Ensslin:99, BB:98, BBK:99, BBK:00}.

The EUV excess in clusters of galaxies was originally attributed to a
diffuse, $(5-10) \times 10^5$ K thermal gas component of the ICM.  Gas
at these temperatures cools rapidly, however, requiring either a
substantial mass of cooling gas or a heretofore unknown energy source
to reheat it \citep{Fabian:96}.  If a reservoir of warm gas were
present in the cores of clusters of galaxies, it would emit strongly in
the far-UV (FUV) resonance lines of \osix\ $\lambda \lambda 1032, 1038$
and \cfour\ $\lambda \lambda 1548, 1551$ as it cools through
temperatures of a few times $10^5$~K \citep{Edgar:Chevalier:86,
Voit:94}, yet \citet{Dixon:Hurwitz:Ferguson:96} were unable to detect
either doublet in the spectra of five clusters of galaxies obtained
with the Hopkins Ultraviolet Telescope (\hut).  To probe more deeply
for this emission, we have observed the Virgo and Coma clusters with
the {\it Far Ultraviolet Spectroscopic Explorer (FUSE)}.  Our limits on
the \osix\ flux of each cluster place tight constraints on the emission
integral of any warm component of their ICM.

\section{OBSERVATIONS AND DATA REDUCTION}

\fuse\ comprises four separate optical systems.  Two employ LiF optical
coatings and are sensitive to wavelengths from 990 to 1187 \AA, while
the other two use SiC coatings, which provide reflectivity to
wavelengths as short as 905 \AA.  The four channels overlap between 990
and 1070 \AA.  For a complete description of \fuse, see \citet{Moos:00}
and \citet{Sahnow:00}.

The \fuse\ spectrum of the Coma cluster was obtained in 17 separate
exposures on 2000 June 18 and 19.  Each was centered on 12\h 59\m 49\fs
0, +27\arcdeg 57\arcmin 46\arcsec\ (J2000, or $l = 57.61, b = +87.96$
in Galactic coordinates), near the cluster center.  The total exposure
was 28608 s, with 23553 s obtained during orbital night.  We use the
entire 29-ksec data set in our analysis.  Our spectrum of the Virgo
cluster is a combination of data from two locations near the cluster
center.  Two exposures, centered on 12\h 31\m 07\fs 3, +12\arcdeg
23\arcmin 46\arcsec\ ($l = 284.03, b = +74.52$) and totaling 2242 s,
were obtained on 2000 June 13.  Seven exposures, centered on 12\h 31\m
13\fs 4, +12\arcdeg 22\arcmin 10\arcsec\ ($l = 284.2, b = +74.50$) and
totaling 8688 s, were obtained on 2000 June 17.  The total integration
time is 10,930 s, all of it during orbital night.  All observations
were made through the 30\arcsec\ $\times$ 30\arcsec\ (LWRS) aperture.

The standard \fuse\ data-reduction software are used to confirm that
none of our data were obtained during passages through the South
Atlantic Anomaly or at low earth-limb angles, to separate day and night
observations, and to confirm that the dead-time correction for both
observations is negligible.  To reduce the detector background, photon
events with pulse heights less than 4 or greater than 15 (in standard
arbitrary units) are excluded from the data set.  Subsequent data
reduction is performed by hand.  Photon events from regions of the
detector with known defects are removed.  The standard extraction
window defined for the LWRS aperture is centered (in the dimension
perpendicular to the dispersion axis) on the diffuse airglow emission
lines.

While the resolution of the \fuse\ spectrograph is approximately 15
\kms\ for a point source,  diffuse emission filling the LWRS aperture
yields a line profile that is well approximated by a top-hat function
with a width of $\sim$ 106 \kms.  Our spectra are quite faint, so we
bin the Coma data by 8 detector pixels, or about 15.5 \kms.  We bin the
data from the shorter Virgo observation by 16 pixels.  We distinguish
between detector pixels and 8- or 16-pixel bins throughout this paper.

We omit the final steps of the standard \fuse\ data-reduction pipeline
for several reasons: first, the pipeline's corrections for differential
Doppler shifts and grating and electronics drifts total less than
10 detector pixels in the dispersion direction.
Second, the distortion correction, applied independently for
each of the instrument apertures, can cause background photons to
``pile up'' in the region between two nominal extraction windows,
forming a horizontal stripe in the processed detector image.  These
stripes occasionally overlap our preferred extraction window.  For
bright point sources, this effect is unimportant, but for faint,
diffuse sources, we prefer not to risk including such features in our
extracted spectrum.  Finally, we wish to include the detector
background as a free parameter in our spectral models, so forgo the
background subtraction steps included in the standard pipeline.

The \fuse\ flux calibration, based on theoretical models of white-dwarf
stellar atmospheres, is believed accurate to about 10\%
\citep{Sahnow:00}.  Corrections to the nominal \fuse\ wavelength scale
are derived from the measured positions of airglow features in each
detector segment and are good to about 0.01 \AA.  In practice, only the
portion of the Coma spectrum obtained during orbital day contains
airglow features strong enough to allow an independent wavelength
correction.  We thus apply the Coma correction to the Virgo data; the
resulting wavelength solution is consistent (within the errors) with
the positions of measureable airglow lines.  Error bars are assigned to
the data assuming Gaussian statistics, then smoothed by 9 bins to
remove small-scale features in the error spectrum without significantly
changing its shape.  Segments of the flux- and wavelength-calibrated
LiF~1A spectra, showing \osix\ $\lambda \lambda 1032, 1038$ at the
redshift of each cluster, are presented in \fig{spectra}.  We assume $z
= 0.0231$ for Coma \citep{Struble:99} and $z = 0.0036$ for Virgo
\citep{Ebeling:98}.  Diffuse emission from our own Galaxy is present in
both spectra; these results are presented in \citet{GalacticOVI}.

\section{SPECTRAL ANALYSIS}

\footnotetext[3]{The Image Reduction and Analysis Facility (IRAF) is
distributed by the National Optical Astronomy Observatories, which is
operated by the Association of Universities for Research in Astronomy,
Inc., (AURA) under cooperative agreement with the National Science
Foundation.}

In our analysis, model spectra are fit to the flux-calibrated data
using the nonlinear curve-fitting program \specfit\ \citep{Kriss:94},
which runs in the IRAF\footnotemark\ environment, to perform a $\chi^2$
minimization of the model parameters.  We find that, because of the
wide disparity in the effective area of the various detector segments,
combining spectra from different segments does not significantly improve
the signal-to-noise ratio.  We thus use data only from the segment with
the highest effective area at the wavelength of interest.

\begin{figure}
\epsscale{0.75}
\plotone{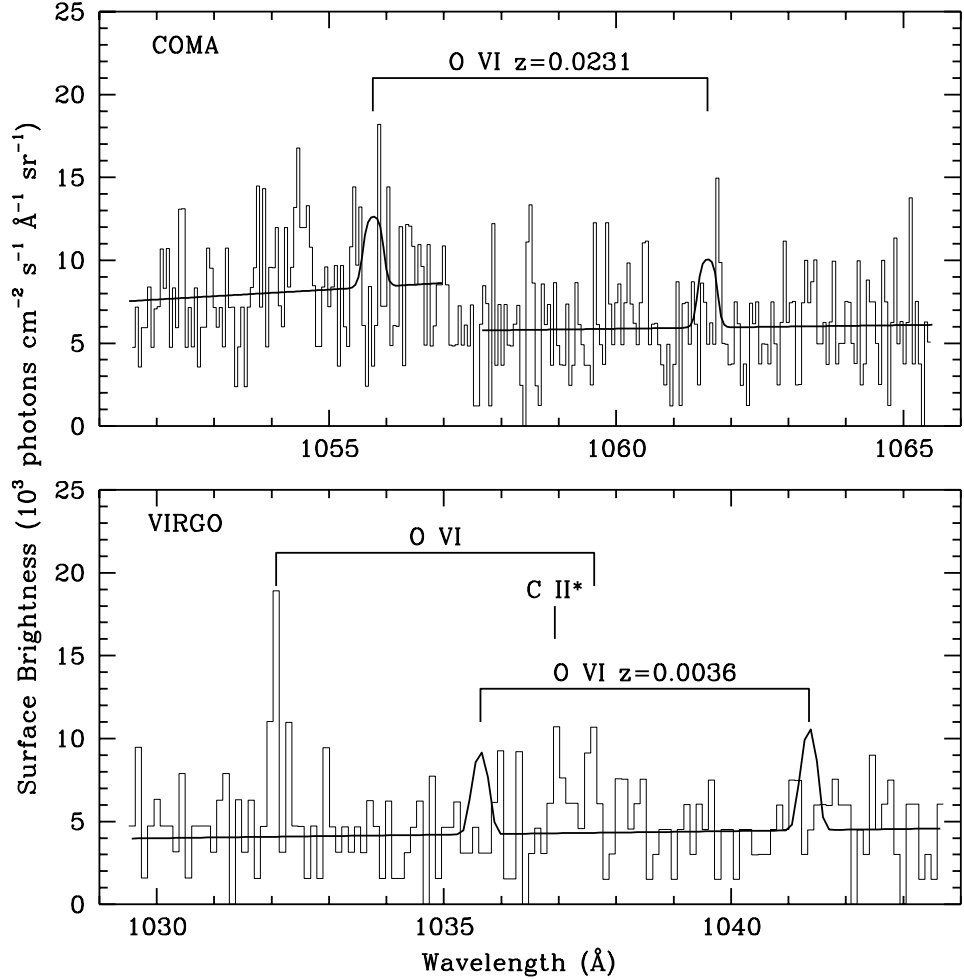}
\caption{\fuse\ spectra of the Coma (top) and Virgo (bottom) clusters,
showing the LiF~1A spectrum in the region about redshifted
\osix\ $\lambda \lambda 1032, 1038$.  The Coma spectrum is binned by 8
detector pixels, the Virgo spectrum by 16.  The data are presented as
histograms and are overplotted by models including \osix\ at our
3-$\sigma$ upper limit.  The observed continuum level is consistent
with the dark-count rate determined from unilluminated regions of the
detector (see text).  Wavelengths between 1045 and 1058 \AA\ are
contaminated by stray light on the LiF~1A detector segment; the
apparent feature at 1054.5~\AA\ in the Coma spectrum is thought to be
an artifact of the stray-light stripe.
\label{spectra}}
\end{figure}

To set limits on the flux of an emission feature, we first fit a linear
continuum to the spectrum.  We then add a synthetic emission feature to
the data array, raising its flux until $\chi^2$ for the best-fit model
with an emission feature differs from $\chi^2$ for the best-fit model
without an emission feature by $\Delta \chi^2 = 9$ \citep[corresponding
to a 3-$\sigma$ deviation for one interesting parameter, in this case
the flux in the emission line;][]{Avni:76}.  We quote as a 3-$\sigma$
upper limit the flux of the model emission line that best fits the
synthetic emission feature originally added to the data.  Upper limits
derived in this way are unaffected by small-scale features in the
observed continua.  The best-fit linear continuum and our 3-$\sigma$
upper limits to the flux of the \osix\ doublet for each cluster are
shown in \fig{spectra}.

The observed profile of a diffuse emission feature represents a
convolution of its intrinsic profile with the 106 \kms\ top-hat function
discussed above.  In our models, we assume the intrinsic profile to be
a Gaussian with FWHM = 40 \kms, corresponding to the thermal width of a
$5 \times 10^5$ K gas.  Broader lines, perhaps due to bulk motions in
the gas, would yield higher upper limits, as the flux is spread over a
greater region of the detector.  For example, lines with an intrinsic
FWHM = 200 \kms\ yield upper limits approximately 70\% higher than those
of 40 \kms\ lines.

No significant FUV line emission is seen at the cluster redshift in
either spectrum.  Our limits on the fluxes of various lines predicted
to be strong in thermal-gas \citep{Landini:90}, mixing-layer
\citep{SSB:93}, and shock \citep{HRH:87} models are presented in Table
\ref{limits}.  Dereddened intensities are calculated assuming the
extinction parameterization of \citet[hereafter CCM]{CCM:89} with
\ebv\ = 0.008 for Coma and 0.030 for Virgo \citep{Schlegel:98} and $R_V
= 3.1$.  Uncertainties in the reddening correction are discussed in
\citet{Dixon:Hurwitz:Ferguson:96}.

At the wavelength of redshifted \osix\ in the Coma cluster (1055.77,
1061.59 \AA), the LiF~1A detector channel has an effective area $A_{\rm
eff} = 24.3$ cm$^{-2}$, by far the highest of the four \fuse\ channels
\citep{Sahnow:00}.  Unfortunately, a bright band of scattered light
contaminates the LiF~1A spectrum from about 1045 to 1058 \AA.  Present
even at night, the band is clearly visible in \fig{spectra}.  It shows
considerable structure, including an apparent emission feature at
1054.5 \AA.  Because the intensity of the band varies on small spatial
scales, background spectra extracted from detector regions above and
below our spectral window cannot be used to predict its shape in our
spectrum.  The statistical significance of the 1054.5 \AA\ feature is
about 3.3 $\sigma$; if real, it should be present in the spectrum from
the less-sensitive ($A_{\rm eff} = 15.5$ cm$^{-2}$) LiF~2B channel at a
significance of $>$ 2 $\sigma$.  No such feature is seen, and we
conclude that it is an artifact of the stray-light stripe in the LiF~1A
detector segment.

Because the \fuse\ SiC channels lie on the sun-illuminated side of the
spacecraft, their spectra may be contaminated by solar emission
\citep{Shelton:01}.  We thus use only data obtained during orbital
night to set limits on emission features in these spectra.  In
practice, only the \ion{S}{6} $\lambda 933.38$ line in our Coma
spectrum is affected, as all other features in the spectrum are
redshifted onto the LiF detector channels.  All of the Virgo data were
obtained during orbital night.

In our Virgo spectrum, the \ion{C}{2}* $\lambda 1037.02$ and \osix\
$\lambda 1037.62$ lines are redshifted to 1040.75 and 1041.36 \AA,
respectively, and, if present, might be contaminated by flux from the
\ion{O}{1} airglow features at 1040.94 and 1041.688 \AA.  To set limits
on the flux of \ion{C}{2}* and \osix, we include the airglow lines in
our model spectrum, but limit their fluxes to that observed in the
\ion{O}{1} $\lambda 1027.43$ line, which is generally much brighter
than either of the two longer-wavelength features.  The resulting
limits are higher than those for the nearby Ly $\beta$ and
\osix\ $\lambda 1031.93$ lines.  Fortunately, all of the Virgo data
were obtained during orbital night, when the \ion{O}{1} line flux is
minimal.

Finally, we set upper limits on the FUV continuum emission from each
cluster of galaxies.  We consider only the region between 1060 and 1070
\AA, which is free of airglow features and for which the LiF~1A
detector has $A_{\rm eff} > 20$ cm$^2$.  For each observation, we
determine the mean background-event rate from unilluminated regions of
the detector directly above and below the LWRS extraction window.
Scaling these rates to the size of the LWRS window, we find that the
observed continuum exceeds the detector background by $110 \pm 250$ and
$200 \pm 400$ \surface, respectively, for Coma and Virgo.  (We ignore
possible structure in the detector background, thought to be present at
the 10\% level.)

\section{DISCUSSION}

\subsection{The Coma Cluster}

From simultaneous fits to \euve\ Deep Survey (DS) and \rosat\ PSPC
observations of the Coma cluster, \citet{Lieu:96a} derive the
temperature and emission integral as a function of radius for a
three-component ICM assuming a distance of 139 Mpc and an abundance $Z
= 0.21$ $Z_{\sun}$ \citep{Hughes:93}.  They find a temperature of $8.7
\times 10^5$~K and an emission integral of $4.8 \times 10^{65}$
cm$^{-3}$ for the coolest component within $3\arcmin$ of the cluster
center.  Combining these parameters with line emissivities estimated
with the CHIANTI software package \citep{CHIANTI} using the
\citet{Feldman:92} solar abundances scaled to the assumed cluster
metallicity, we derive an expected \osix\ $\lambda 1032$ flux of $3.8
\times 10^{-3}$ photons cm$^{-2}$ s$^{-1}$.  If the \osix\ emission
were evenly distributed within this region (an assumption consistent
with both EUV and soft X-ray maps of the cluster;
\citealt{Bonamente:00, Briel:92}), then its surface brightness would be
1600 photons cm$^{-2}$ s$^{-1}$ sr$^{-1}$, slightly less than our
3-$\sigma$ upper limit to the dereddened \osix\ surface brightness.

\begin{figure}
\plotone{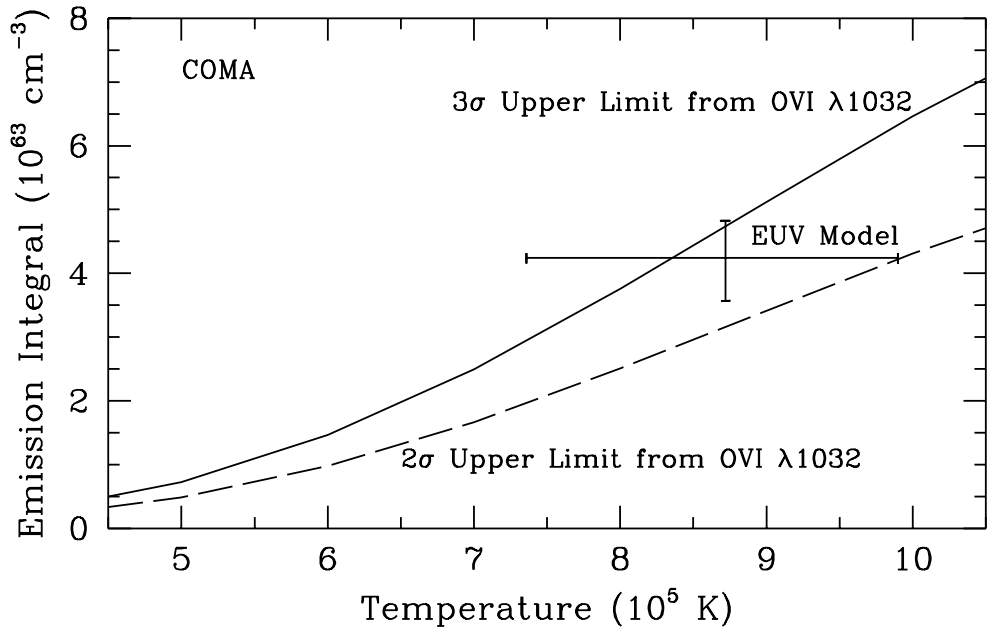}
\caption{Upper limits to the emission integral ($EI \equiv \int n_p n_e
dV$) as a function of gas temperature in the Coma cluster.  The solid
curve is derived from the dereddened 3-$\sigma$ upper limit to the
\osix\ $\lambda 1032$ surface brightness, assuming the distance (139
Mpc) and abundance ($Z = 0.21$ $Z_{\sun}$) of \citealt{Lieu:96a}.  The
dashed line represents a 2-$\sigma$ limit to the same feature.  The
data point indicates the best-fit value of the emission integral and
temperature derived for the central region ($r < 3\arcmin$) of the
cluster by \citealt{Lieu:96a}, scaled to the area of the \fuse\ LWRS
aperture.  The \fuse\ data nominally exclude the warm-gas model at the
2-sigma level.
\label{ei_coma}}
\end{figure}

The \osix\ line emissivity depends critically on the gas temperature.
From our limit to the \osix\ flux of Coma, we derive an upper limit to
the emission integral as a function of temperature. In \fig{ei_coma},
we compare this limit with the best-fit value of the emission integral
derived by \citet{Lieu:96a} for the central region ($r < 3\arcmin$) of
the cluster, scaled to the area of the LWRS slit.  This value of the
emission integral, $EI = 4.2 \times 10^{63}$ cm$^{-3}$, lies just below
the \fuse\ 3-$\sigma$ upper limit.  If we instead use a 2-$\sigma$
limit to the \osix\ flux, we can nominally exclude the warm-gas model
of \citet{Lieu:96a}.  More recent models of the Coma EUV excess do not
require a sub-$10^6$~K gas.  \citet{Bonamente:00} fits the data with a
two-component ICM, the cooler of which has a temperature of $2.9 \times
10^6$ K.  We would not expect significant \osix\ emission from such a
hot gas.

\subsection{The Virgo Cluster}

Two groups \citep{Lieu:96b, BLM:01} have modeled the EUV excess
in the Virgo cluster with a warm component of the ICM.  Both find that
its temperature rises with radius, while its emission integral falls.
The brightest FUV emission would thus be expected from the cluster
center.  Unfortunately, the center of the Virgo cluster is occupied by
the galaxy M87, and its strong stellar continuum complicates the FUV
spectrum of this region \citep[\cf,][]{Dixon:Hurwitz:Ferguson:96}.  We
thus consider data taken at larger radii, about 4\farcm 4 (2242 s)
and 6\farcm 0 (8688 s) from the cluster center.
Both \citet{Lieu:96b} and \citet{BLM:01} model the warm ICM using a series
of concentric annuli.  Our inner data point samples the 3--5 arcmin
annulus, while our outer point falls in the 5--7 arcmin region.
Because 80\% of our data comes from the outer annulus, we compare our
observations with model predictions for the 5--7 arcmin annulus.  We
estimate that errors in our derived limits to the emission integral
resulting from this simplification are on the order of the
uncertainties in the \fuse\ flux calibration.

\begin{figure}
\plotone{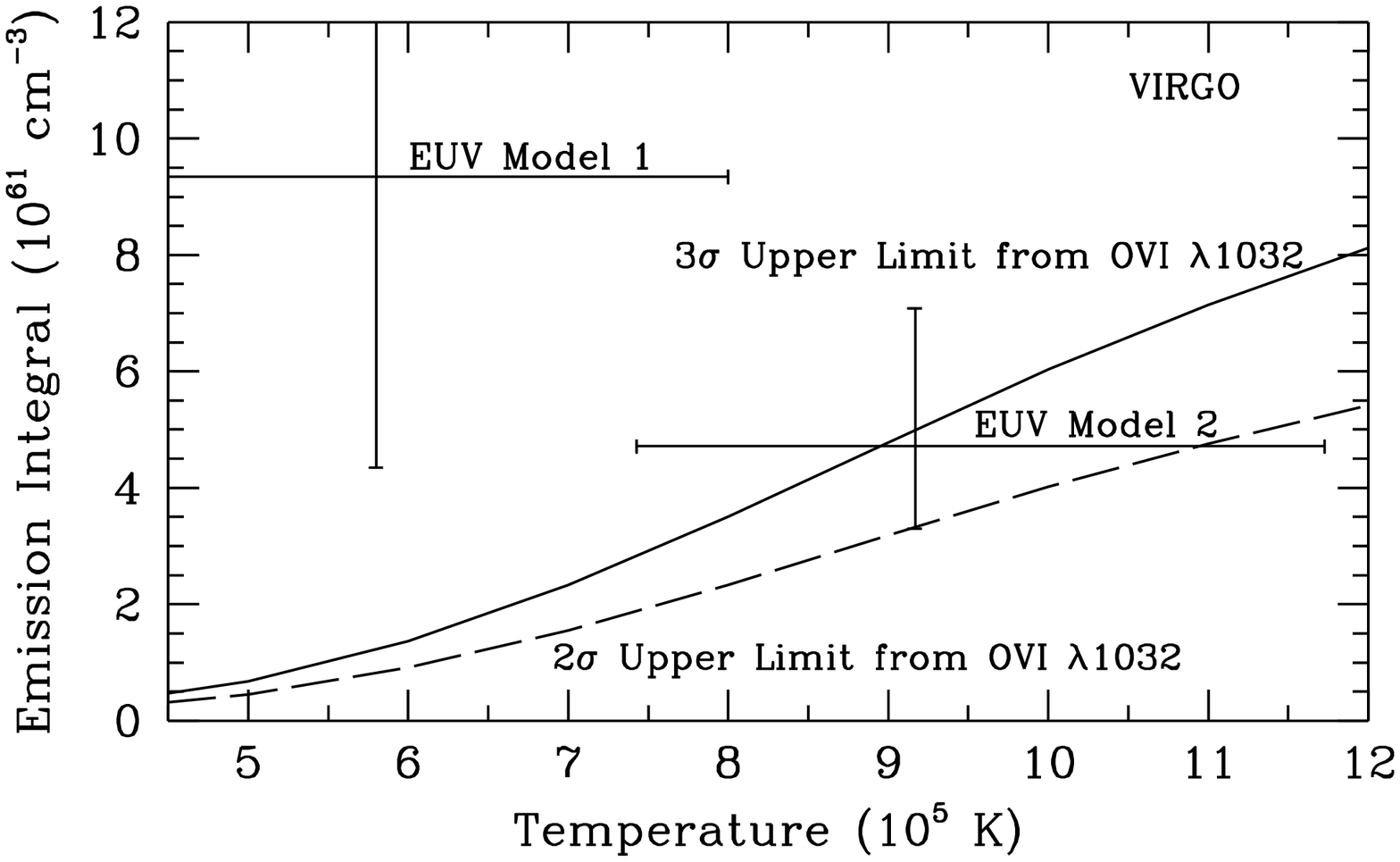}
\caption{Same as \fig{ei_coma}, but for Virgo.  Our limit to the
emission integral assumes the distance (17.2 Mpc) and abundance ($Z =
0.5$ $Z_{\sun}$) used by \citealt{BLM:01}.  The data points indicate
the best-fit value of the emission integral and temperature derived for
the 5--7 arcmin annulus by \citealt{Lieu:96b} (EUV Model 1) and
\citealt{BLM:01} (EUV Model 2), scaled to the area of the \fuse\ LWRS
aperture and our assumed distance.  Model 1 is completely excluded by
our limit, while Model 2 is nominally excluded at the 2-sigma level.
\label{ei_virgo}}
\end{figure}

Our upper limit to the emission integral as a function of temperature
is plotted in \fig{ei_virgo}.  Also shown are the best-fit values of
the temperature and emission integral of the warm ICM, derived from
simultaneous fits to the DS and PSPC data from the 5--7 arcmin annulus
of the Virgo cluster by \citet{Lieu:96b} and \citet{BLM:01},
respectively, scaled to the area of the LWRS slit.  The Lieu
\etal\ model parameters are $T = 5.8 \times 10^5$ K, $EI = 9.35 \times
10^{61}$ cm$^{-3}$, and $Z = 0.454$ $Z_{\sun}$, values firmly excluded
by our limit to the \osix\ flux.  The Bonamente \etal\ model employs a
much warmer gas, with $T = 9.2 \times 10^5$ K, $EI = 4.72 \times
10^{61}$ cm$^{-3}$, and $Z = 0.5$ $Z_{\sun}$; these parameters fall
just below the \fuse\ 3-$\sigma$ upper limit.  If we instead use a
2-$\sigma$ limit to the \osix\ flux, we can nominally exclude this
model as well.

The redshift of M87 ($z = 0.00436$; \citealt{Smith:00}) is
slightly greater than that of the Virgo cluster mean.  Near the center
of the cluster, the X-ray emission is well centered on M87, and certain
asymmetrical features seem correlated with radio emission from the
nucleus \citep{Belsole:01}.  If, in this region, the cooling flow emits
at the redshift M87, the \osix\ $\lambda 1032$ line would be strongly
attenuated by absorption from Galactic \ctwo\ $\lambda 1036.3$.  An
upper limit to the emission integral derived from the \osix\ $\lambda
1038$ line would be approximately twice our quoted value.

\section{CONCLUSIONS}

We have obtained FUV spectra of two clusters of galaxies with \fuse,
the {\it Far Ultraviolet Spectroscopic Explorer}.  Neither spectrum
shows significant \osix\ $\lambda \lambda 1032, 1038$ emission at the
cluster redshift.  Our 2-$\sigma$ upper limits on the \osix\ $\lambda
1032$ flux from Coma and Virgo exclude all published warm-gas models of
the EUV excess in these clusters and severely constrain the amount of
gas at $T \lesssim 8 \times 10^5$ K in either cluster.

\acknowledgments

We thank R. Shelton and E. Murphy for discussions of \fuse\ data
analysis and D. Buote for thoughtful comments on the manuscript.  We
acknowledge the outstanding efforts of the \fuse\ P.I.\ team to make
this mission successful.  This research has made use of NASA's
Astrophysics Data System and the NASA/IPAC Extragalactic Database
(NED), operated by JPL for NASA.  This work is supported by NASA grant
NAG 5-5197.



\clearpage

\begin{deluxetable}{lccccc}
\tablewidth{0pt}
\tablecaption{3-$\sigma$ Upper Limits to Far-UV Emission Lines \label{limits}}
\tablecolumns{6}
\tablehead{
& \colhead{Rest Wavelength} &
\multicolumn{2}{c}{Coma Limits} &
\multicolumn{2}{c}{Virgo Limits} \\
\colhead{Species}  & \colhead{(\AA)}  &
\colhead{Observed} & \colhead{Dereddened} &
\colhead{Observed} & \colhead{Dereddened}
}
\startdata
\ion{S}{6}  & \phn933.38 & 3200\tablenotemark{a} & 3700 & 4400 & 7600 \\
\ion{C}{3}  & \phn977.02 & 3900 & 4400 & 4800 & 7700 \\
\ion{N}{3}  & \phn989.80 & 1600 & 1800 & 4700 & 7400 \\
\ion{Ne}{6}] & \phn999.20 & 1300 & 1500 & 4500 & 7000 \\
\ion{H}{1} (Ly $\beta$) & 1025.72 & 1300 & 1400 & 2000 & 3000 \\
\osix\      & 1031.93    & 1600 & 1800 & 1700 & 2600 \\
\ion{C}{2}* & 1037.02    & 1200 & 1300 & 2400 & 3600 \\
\osix\      & 1037.62    & 1500 & 1700 & 2100 & 3200 \\
\ion{Ne}{5}] & 1146.1\phn & 2200 & 2400 & 2900 & 4000
\enddata

\tablenotetext{a}{To minimize contamination of the SiC~2A channel by
solar emission, this upper limit is derived only from Coma data taken
during orbital night.  (All of the Virgo data were taken during orbital
night.)}

\tablecomments{Units are photons cm$^{-2}$ s$^{-1}$ sr$^{-1}$.  Upper
limits are derived assuming an intrinsic line width of 40 \kms\ (FWHM)
at the cluster redshift.  Dereddened limits assume a CCM extinction
curve with \ebv\ = 0.008 for Coma and 0.030 for Virgo and $R_V = 3.1$.
}
\end{deluxetable}


\newpage

\begin{thebibliography}{38}
\expandafter\ifx\csname natexlab\endcsname\relax\def\natexlab#1{#1}\fi

\bibitem[{{Arabadjis} \& {Bregman}(1999)}]{Arabadjis:99}
{Arabadjis}, J.~S. \& {Bregman}, J.~N. 1999, \apj, 514, 607

\bibitem[{{Avni}(1976)}]{Avni:76}
{Avni}, Y. 1976, \apj, 210, 642

\bibitem[{{Belsole} {et~al.}(2001)}]{Belsole:01}
{Belsole}, E. {et~al.} 2001, \aap, 365, L188

\bibitem[{{Bergh{\"o}fer} {et~al.}(2000){Bergh{\"o}fer}, {Bowyer}, \&
  {Korpela}}]{BBK:00}
{Bergh{\"o}fer}, T.~W., {Bowyer}, S., \& {Korpela}, E. 2000, \apj, 535, 615

\bibitem[{{Bonamente}(2000)}]{Bonamente:00}
{Bonamente}, M. 2000, PhD thesis, University of Alabama in Huntsville

\bibitem[{{Bonamente} {et~al.}(2001){Bonamente}, {Lieu}, \& {Mittaz}}]{BLM:01}
{Bonamente}, M., {Lieu}, R., \& {Mittaz}, J. P.~D. 2001, \apj, in press
  \verb!(astro-ph/0011186)!

\bibitem[{{Bowyer} \& {Bergh{\"o}fer}(1998)}]{BB:98}
{Bowyer}, S. \& {Bergh{\"o}fer}, T.~W. 1998, \apj, 506, 502

\bibitem[{{Bowyer} {et~al.}(1999){Bowyer}, {Bergh{\"o}fer}, \&
  {Korpela}}]{BBK:99}
{Bowyer}, S., {Bergh{\"o}fer}, T.~W., \& {Korpela}, E.~J. 1999, \apj, 526, 592

\bibitem[{{Bowyer} {et~al.}(1997){Bowyer}, {Lieu}, \& {Mittaz}}]{BLM:97}
{Bowyer}, S., {Lieu}, R., \& {Mittaz}, J. 1997, in IAU Symp. 188, The Hot
  Universe, ed. K.~{Koyama}, S.~{Kitamoto}, \& M.~{Itoh} (Dordrecht: Kluwer),
  52

\bibitem[{{Briel} {et~al.}(1992){Briel}, {Henry}, \&
  {B\"{o}hringer}}]{Briel:92}
{Briel}, U.~G., {Henry}, J.~P., \& {B\"{o}hringer}, H. 1992, \aap, 259, L31

\bibitem[{{Cardelli} {et~al.}(1989){Cardelli}, {Clayton}, \& {Mathis}}]{CCM:89}
{Cardelli}, J.~A., {Clayton}, G.~C., \& {Mathis}, J.~S. 1989, \apj, 345, 245

\bibitem[{{Dere} {et~al.}(1997){Dere}, {Landi}, {Mason}, {Monsignori Fossi}, \&
  {Young}}]{CHIANTI}
{Dere}, K.~P., {Landi}, E., {Mason}, H.~E., {Monsignori Fossi}, B.~C., \&
  {Young}, P.~R. 1997, \aaps, 125, 149

\bibitem[{{Dixon} {et~al.}(1996){Dixon}, {Hurwitz}, \&
  {Ferguson}}]{Dixon:Hurwitz:Ferguson:96}
{Dixon}, W.~V., {Hurwitz}, M., \& {Ferguson}, H.~C. 1996, \apjl, 469, L77

\bibitem[{{Dixon} {et~al.}(2001){Dixon}, {Sallmen}, {Hurwitz}, \&
  {Lieu}}]{GalacticOVI}
{Dixon}, W.~V., {Sallmen}, S., {Hurwitz}, M., \& {Lieu}, R. 2001, \apjl,
  submitted

\bibitem[{{Drake}(1999)}]{Drake:99}
{Drake}, J.~J. 1999, \apjs, 122, 269

\bibitem[{{Ebeling} {et~al.}(1998){Ebeling}, {Edge}, {Bohringer}, {Allen},
  {Crawford}, {Fabian}, {Voges}, \& {Huchra}}]{Ebeling:98}
{Ebeling}, H., {Edge}, A.~C., {Bohringer}, H., {Allen}, S.~W., {Crawford},
  C.~S., {Fabian}, A.~C., {Voges}, W., \& {Huchra}, J.~P. 1998, \mnras, 301,
  881

\bibitem[{{Edgar} \& {Chevalier}(1986)}]{Edgar:Chevalier:86}
{Edgar}, R.~J. \& {Chevalier}, R.~A. 1986, \apjl, 310, L27

\bibitem[{{En{\ss}lin} {et~al.}(1999){En{\ss}lin}, {Lieu}, \&
  {Biermann}}]{Ensslin:99}
{En{\ss}lin}, T.~A., {Lieu}, R., \& {Biermann}, P.~L. 1999, \aap, 344, 409

\bibitem[{{Fabian}(1996)}]{Fabian:96}
{Fabian}, A.~C. 1996, Science, 271, 1244

\bibitem[{{Feldman} {et~al.}(1992){Feldman}, {Mandelbaum}, {Seely}, {Doschek},
  \& {Gursky}}]{Feldman:92}
{Feldman}, U., {Mandelbaum}, P., {Seely}, J.~F., {Doschek}, G.~A., \& {Gursky},
  H. 1992, \apjs, 81, 387

\bibitem[{{Forman} \& {Jones}(1982)}]{Forman:Jones:82}
{Forman}, W. \& {Jones}, C. 1982, \araa, 20, 547

\bibitem[{{Hartigan} {et~al.}(1987){Hartigan}, {Raymond}, \&
  {Hartmann}}]{HRH:87}
{Hartigan}, P., {Raymond}, J., \& {Hartmann}, L. 1987, \apj, 316, 323

\bibitem[{{Hughes} {et~al.}(1993){Hughes}, {Butcher}, {Stewart}, \&
  {Tanaka}}]{Hughes:93}
{Hughes}, J.~P., {Butcher}, J.~A., {Stewart}, G.~C., \& {Tanaka}, Y. 1993,
  \apj, 404, 611

\bibitem[{{Krick} {et~al.}(2000){Krick}, {Arabadjis}, \& {Bregman}}]{Krick:00}
{Krick}, J., {Arabadjis}, J.~S., \& {Bregman}, J.~N. 2000, \baas, 32, 1403

\bibitem[{{Kriss}(1994)}]{Kriss:94}
{Kriss}, G.~A. 1994, in ASP Conf. Ser. 61, Astronomical Data Analysis Software
  and Systems III, ed. D.~R. {Crabtree}, R.~J. {Hanisch}, \& J.~{Barnes} (San
  Francisco: ASP), 437

\bibitem[{{Landini} \& {Monsignori Fossi}(1990)}]{Landini:90}
{Landini}, M. \& {Monsignori Fossi}, B.~C. 1990, \aaps, 82, 229

\bibitem[{{Lieu} {et~al.}(1999){Lieu}, {Ip}, {Axford}, \&
  {Bonamente}}]{LIAB:99}
{Lieu}, R., {Ip}, W.-H., {Axford}, W.~I., \& {Bonamente}, M. 1999, \apjl, 510,
  L25

\bibitem[{{Lieu} {et~al.}(1996{\natexlab{a}}){Lieu}, {Mittaz}, {Bowyer},
  {Breen}, {Lockman}, {Murphy}, \& {Hwang}}]{Lieu:96a}
{Lieu}, R., {Mittaz}, J. P.~D., {Bowyer}, S., {Breen}, J.~O., {Lockman}, F.~J.,
  {Murphy}, E.~M., \& {Hwang}, C.-Y. 1996{\natexlab{a}}, Science, 274, 1335

\bibitem[{{Lieu} {et~al.}(1996{\natexlab{b}}){Lieu}, {Mittaz}, {Bowyer},
  {Lockman}, {Hwang}, \& {Schmitt}}]{Lieu:96b}
{Lieu}, R., {Mittaz}, J. P.~D., {Bowyer}, S., {Lockman}, F.~J., {Hwang}, C.-Y.,
  \& {Schmitt}, J. H. M.~M. 1996{\natexlab{b}}, \apjl, 458, L5

\bibitem[{{Mittaz} {et~al.}(1998){Mittaz}, {Lieu}, \& {Lockman}}]{MLL:98}
{Mittaz}, J. P.~D., {Lieu}, R., \& {Lockman}, F.~J. 1998, \apjl, 498, L17

\bibitem[{{Moos} {et~al.}(2000)}]{Moos:00}
{Moos}, H.~W. {et~al.} 2000, \apjl, 538, L1

\bibitem[{{Sahnow} {et~al.}(2000)}]{Sahnow:00}
{Sahnow}, D.~J. {et~al.} 2000, \apjl, 538, L7

\bibitem[{{Schlegel} {et~al.}(1998){Schlegel}, {Finkbeiner}, \&
  {Davis}}]{Schlegel:98}
{Schlegel}, D.~J., {Finkbeiner}, D.~P., \& {Davis}, M. 1998, \apj, 500, 525

\bibitem[{{Shelton} {et~al.}(2001)}]{Shelton:01}
{Shelton}, R.~L. {et~al.} 2001, \apj, submitted

\bibitem[{{Slavin} {et~al.}(1993){Slavin}, {Shull}, \& {Begelman}}]{SSB:93}
{Slavin}, J.~D., {Shull}, J.~M., \& {Begelman}, M.~C. 1993, \apj, 407, 83

\bibitem[{{Smith} {et~al.}(2000){Smith}, {Lucey}, {Hudson}, {Schlegel}, \&
  {Davies}}]{Smith:00}
{Smith}, R.~J., {Lucey}, J.~R., {Hudson}, M.~J., {Schlegel}, D.~J., \&
  {Davies}, R.~L. 2000, \mnras, 313, 469

\bibitem[{{Struble} \& {Rood}(1999)}]{Struble:99}
{Struble}, M.~F. \& {Rood}, H.~J. 1999, \apjs, 125, 35

\bibitem[{{Voit} {et~al.}(1994){Voit}, {Donahue}, \& {Slavin}}]{Voit:94}
{Voit}, G.~M., {Donahue}, M., \& {Slavin}, J.~D. 1994, \apjs, 95, 87

\end{thebibliography}
\end{document}